\documentclass[]{JHEP3} 

\usepackage{amsmath}
\usepackage{amscd}
\usepackage{amsfonts}
\usepackage{amsthm}

\newcommand{\be}{\begin{equation}}
\newcommand{\ee}{\end{equation}}
\newcommand{\cF}{{\cal F}}
\newcommand{\pol}{\mathrm{pol}}
\newcommand{\sla}{\slash\!\!\!}
\newcommand{\dual}[1]{{}^\star#1}
\newcommand{\Tr}{\mathrm{Tr} \, }
\newcommand{\dagg}{{}^\dagger}
\newcommand{\Ann}{\mathrm{Ann}\,}

\title{Stable D-branes, calibrations and generalized Calabi-Yau geometry}

\author{Paul Koerber\\
    Department of Physics and Astronomy, University of British Columbia\\
    6224 Agricultural Road,Vancouver, B.C., V6T 1Z1, Canada\\
    E-mail: {\tt koerber} at {\tt physics.ubc.ca}}

\preprint{\hepth{0506154}}

\abstract{We introduce generalized calibrations that take into account the gauge
field on the D-brane so that calibrated submanifolds minimize the Dirac-Born-Infeld energy.
We establish the calibration bound and show that the calibration form is closed in a supersymmetric
background with non-vanishing NS-NS $3$-form $H$ and dilaton $\Phi$. We show that the calibration conditions
are equivalent to the existence of unbroken supersymmetry on the D-brane. We study the problem of
supersymmetric D-branes in the presence of $H\neq 0$ also from the world-sheet approach and find exactly
the same conditions. Finally, we show that our notion of generalized calibrations is equivalent to the
calibrations introduced in the context of generalized Calabi-Yau geometry in {\tt math.DG/0401221}.}


\begin{document}


\section{Introduction}

String/M-theory on supersymmetric backgrounds with non-vanishing fluxes is currently a very active
field of study. One reason is that those backgrounds provide the setup for models with attractive
phenomenology and another is that they appear in generalizations of the AdS/CFT correspondence.
The background geometry in this paper consists of non-vanishing fields in the common NS-NS
sector of type IIA and IIB supergravities, i.e.\ we consider a non-vanishing dilaton $\Phi$
and 3-form $H$, but put all R-R fields and fermions to zero. The supersymmetry conditions for backgrounds
with fluxes, pioneered in \cite{strominger}, lead to $G$-structures. We will mainly consider geometries
with $SU(n)_L \times SU(n)_R$-structure, where the $SU(n)_{L/R}$ are constant with respect
to covariant derivatives with different connections $\nabla \pm \frac{1}{2}H$ \cite{bihermitian}.

In this paper we are interested in the conditions for branes to preserve some of the
supersymmetry of the background. In the simplest case, without fluxes, the background has special
holonomy and supersymmetric branes wrap calibrated submanifolds \cite{harveylawson},
which are volume-minimizing \cite{becker1,becker2,gibbonscal}. For $SU(n)$ holonomy (Calabi-Yau) there
are two cases depending on whether the calibration is $e^{i \omega}$ or $\Re(\Omega)$, where $\omega$
is the K\"ahler form and $\Omega$ is the $(n,0)$-form. These correspond to complex and special Lagrangian
submanifolds respectively.

In supersymmetric backgrounds with fluxes, supersymmetric branes are associated with generalized calibrations,
which were introduced in \cite{gencal} and extensively studied in \cite{gencal2,gencal3}. These calibrations
take into account the coupling of branes with background fluxes so that the calibrated submanifolds are
no longer volume-minimizing but rather energy-minimizing. Here we introduce another notion of generalized
calibrations, in the same general philosophy though, which takes into account the gauge field $\cF$ on a D-brane.
As far as the author is aware a
calibration like this has not yet been introduced for general D$p$-branes (see \cite{fivebrane} for a brief
discussion of the case of the D4-brane as dimensional reduction of the M5-brane). Generalized calibrations
now minimize the Dirac-Born-Infeld energy. Furthermore it is shown that the calibration conditions
are equivalent to the vanishing of the gluino supersymmetry transformation for some spinors. The conditions
for the latter were studied in \cite{minasian}.

However, we can study these conditions also from the string world-sheet viewpoint where
D-branes are regarded as boundary conditions for open strings.
In the case of vanishing 3-form flux and flat gauge field on the D-brane, $\cF=0$, it is well-known that this approach gives equivalent
results \cite{ooguri,becker2}. The string world-sheet approach starts from an $N=(2,2)$ SCFT
in the bulk, which induces $U(n)_L \times U(n)_R$ structure, and demands that the boundary conditions
preserve $N=2$ world-sheet supersymmetry. This is precisely the condition for the D-branes to descend
to topological string theory so they are called topological branes. Depending on which combination
of left- and right-moving supersymmetry is preserved one
has B-type and A-type D-branes corresponding to the complex and special Lagrangian submanifolds of the
effective action approach respectively.

In \cite{kapustincoisotropic} it was discovered that there exist supersymmetric D-branes of type A
which are not special Lagrangian if the gauge field $\cF$ is turned on. In that paper the condition
for the D-branes to be topological was worked out: they are coisotropic rather than Lagrangian. However,
the requirement of $N=2$ world-sheet supersymmetry alone is not enough for target space supersymmetric
D-branes. To proceed one should note that target space supersymmetry is generated by the spectral flow
operators. In order to globally define these spectral flow operators and thus have preserved target space
supersymmetry in the bulk we must further reduce the structure to $SU(n)_L \times SU(n)_R$. On the boundary, one
needs preservation of the spectral flow operator, which is called the
stability condition. In the simplest case of $\cF=0$ stability corresponds to the requirement of {\em special}
Lagrangian in addition to just Lagrangian. In \cite{kapustinstable} this stability condition was studied in the case
of non-vanishing gauge field on a D-brane in a Calabi-Yau manifold ($H=0$) and shown to be completely equivalent to
the conditions for supersymmetric
D-branes found from the effective action approach in \cite{minasian}.

In this paper we generalize the world-sheet approach to the case $H \neq 0$. The topological string theory
with $H \neq 0$ was introduced in \cite{kapustintopological} and the condition for the D-brane
to be topological was studied in \cite{zabzineboundary}. Here we construct the remaining condition for the
D-brane to be stable and show that both requirements, topological and stable, are exactly the same as the
conditions for the D-brane to be generalized calibrated. Therefore, also in the case $H \neq 0$
we find the same supersymmetry requirements from the world-sheet approach as from the effective action approach.

A geometry with $U(n)_L \times U(n)_R$-structure where the $U(n)$ structures are covariantly constant with
respect to different connections $\nabla \pm \frac{1}{2} H$, called bihermitian geometry \cite{bihermitian}, is
in fact completely equivalent to a (twisted) generalized K\"ahler structure, introduced in \cite{gualtieri},
building on the work of \cite{hitchin}. A generalized K\"ahler structure consists of two commuting (twisted)
generalized complex structures $(\mathcal{J}_1,\mathcal{J}_2)$. If the structure is further reduced to
$SU(n)_L \times SU(n)_R$ we call it a generalized Calabi-Yau geometry\footnote{Note that the definition
of a generalized Calabi-Yau structure in \cite{hitchin} is different from that in \cite{gualtieri}. Here we mean the
latter stronger one.}. It comes as no surprise then that our generalized calibrations should have an interpretation
in the theory of generalized complex structures.
In \cite{zabzineboundary} it was shown that a D-brane is topological if it is a generalized complex
submanifold with respect to $\mathcal{J}_1$ (for type B) or $\mathcal{J}_2$ (for type A). Furthermore, in
\cite{gualtieri} a definition of a calibration in a generalized Calabi-Yau geometry was given. We show that
our concept of generalized calibration is equivalent i.e.\ a brane is generalized calibrated (in the sense
this paper) if it is a generalized complex submanifold with respect to $\mathcal{J}_1$ and
calibrated (in the sense of \cite{gualtieri}) with respect to $\mathcal{J}_2$ for type B and vice-versa for type A.
Under the mirror symmetry
automorphism of the world-sheet theory $\mathcal{J}_1$ and $\mathcal{J}_2$ are exchanged so that mirror symmetry
indeed swaps B- and A-type branes. Furthermore, we note that B-type topological string theory defined in
\cite{kapustintopological} only sees $\mathcal{J}_1$ while the stability condition of the B-brane depends on
$\mathcal{J}_2$ and vice-versa for the A-brane. This is in fact also the case for $H=0$, where the roles
of complex structure and K\"ahler form are exchanged between the topological field theory dependence and the
stability criteria.

Other work on generalized complex structures from the target space viewpoint is \cite{grana1,grana2,grana3},
from the world-sheet viewpoint \cite{worldsheet1,worldsheet2,worldsheet3} and on the relation with mirror
symmetry \cite{ms1,ms2,ms3,ms4}.

In section \ref{cal} a definition of generalized calibrations is given. The calibrated submanifolds minimize
the Dirac-Born-Infeld energy. A suitable calibration form is constructed from the generators of unbroken
supersymmetry. It is shown that this form is closed and generates the calibration bound. We show that the
conditions for saturating the bound coincide with the condition for supersymmetric cycles. In section \ref{worldsheet}
the same conditions for supersymmetry are found, but now from the world-sheet approach.  In section \ref{CY} we present
the canonical example of ordinary Calabi-Yau manifolds. In section \ref{gencomp}
the results are interpreted in the context of generalized Calabi-Yau geometry.

\section{Calibrations}
\label{cal}
\subsection{Generalized calibrations}

In this subsection we will quickly review the concept of calibrations
and generalize it slightly to include the gauge field living on the world-volume
of D-branes. Calibrations were introduced in \cite{harveylawson} in order to construct
volume minimizing submanifolds.

An oriented tangent $p$-plane is a vector subspace $V$ of $T_x M$
with an orientation. A $p$-form $\phi$ on a Riemannian manifold $(M,g)$ is a
{\em calibration} if $d\phi=0$ ($\phi$ is closed) and for any tangent $p$-plane, $V$,
it satisfies
\be
\phi|_V \le \mathrm{vol}|_V,
\ee
where $\phi|_V$ is the pull-back to $V$ and $\mathrm{vol}|_V$ is the induced volume form on $V$. We
also demand that in every point $x$ of $M$, there exist $p$-planes for which the bound is saturated. Those $p$-planes
form the {\em contact set}.

A $p$-dimensional submanifold $N$ of $M$ --- a $p$-brane --- is calibrated by $\phi$ if
at any point $x \in N$
it satisfies $\phi|_{T_x N}=\mathrm{vol}|_{T_x N}$ i.e.\ it saturates the bound at any point $x$.
In this paper we will often rewrite this condition as
\be
P(\phi)_\epsilon = \sqrt{P(g)},
\ee
where $P$ denotes the pullback to the $p$-brane world-volume and\footnote{See appendix \ref{conventions} for more conventions.}
\be
P(\phi)_\epsilon = \frac{1}{p!} P(\phi)_{a_1\ldots a_p} \epsilon^{a_1\ldots a_p}.
\ee
It is clear that such branes are world-volume minimizing within their homology class
since if we take another brane $N'$ within the same class $N'=N + \partial Q$ we find
\be
\mathrm{vol}(N') = \int_{N'} \mathrm{vol} \ge \int_{N'} \phi = \int_N \phi + \int_Q d \phi = \int_N \phi = \int_N \mathrm{vol}=\mathrm{vol}(N),
\label{minimal}
\ee
where we used Stokes' theorem and $d\phi=0$. Calibrations are often constructed from bilinears in spinors \cite{calibrationbook}.
One can then make the link with supersymmetry generators and show that calibrated branes preserve some of the supersymmetry of the background.

In \cite{gencal} generalized calibrated submanifolds were introduced which do not minimize the
volume but rather the brane energy, which includes couplings to the background fields. Likewise these branes
wrap supersymmetric cycles. In this paper, however, we specialize to D-branes and also take into account the
gauge field $\cF$, with $d \cF=0$, on the D-brane. The basic philosophy of generalized calibrations is the same in
that we will now minimize the D-brane energy i.e.\ the Dirac-Born-Infeld energy. A D-brane is now a
generalized submanifold with data $(N,\cF)$ with $\cF$ an abelian gauge field.
We introduce a sum of forms of different dimension $\phi \in \wedge^\bullet T_M^*$,
\be
\phi= \sum_l \phi^{(l)},
\ee
and a polynomial in $\cF$, $\pol(\cF)$, in which the products are wedges. $\phi$ is a {\em generalized
calibration} if $d\phi=0$ and for every generalized submanifold $(N,\cF)$ the following bound is satisfied:
\be
\left(P(\phi) \wedge \pol(\cF)\right)_{[p],\epsilon} \le e^{-\Phi} \sqrt{P(g-b)+\cF},
\label{bound}
\ee
where we selected out the $p$-dimensional part of $P(\phi) \wedge \pol(\cF)$, $b$ is the NS-NS field and $\Phi$
the dilaton.
It will be convenient to introduce $F=\cF-P(b)$, since $\cF$ and $P(b)$ will always appear in this combination in D-brane actions.
As usual the torsion is given by $H=db$. The right-hand side of the bound is indeed the Dirac-Born-Infeld energy.
We had to go to the trouble of considering $\phi$ separately from $\cF$ because the form $\phi$ is defined on $M$ while $\cF$
is only defined on the D-brane world-volume and in fact part of the D-brane data. The reader should keep in mind that this concept
of generalized calibrations is different from \cite{gencal}. We will use the term generalized calibrations in the hope that
it will cause no confusion.
Now $(N,\cF)$ is a {\em generalized calibrated submanifold} if
\be
\left(P(\phi) \wedge \pol(\cF)\right)_\epsilon = e^{-\Phi} \sqrt{P(g)+F}.
\ee
If we now take another D-brane $(N',\cF')$ where $N'=N + \partial Q$ is in the same homology class as $N$
we can go through the same reasoning as in \eqref{minimal} to show that $(N,\cF)$ has indeed minimal energy within its class
provided that
\be
\int P_{N'}(\phi) \wedge \pol(\cF') = \int P_N(\phi) \wedge \pol(\cF).
\ee
The exact topological condition for this statement to be true is that there must exist a gauge bundle
on $Q$ such that its Chern class restricted to $N$ gives the Chern class of $\cF$ and its Chern class restricted
$N'$ the Chern class of $\cF'$. It might be better to choose a gauge bundle on the whole manifold $M$ right
from the start (the choice of a particular gauge field $\cF$ within that bundle is still free as it should be since it is part
of the data of the D-brane) although we loose some generality then\footnote{We thank Jim Bryan for explaining all this.}.

\subsection{The calibration form}

In this subsection we construct the calibration form $\phi$ and show that it is closed while in the next subsection
we will prove the bound \eqref{bound}.
The basic ingredients of our calibration form are the generators of left- and
right-moving preserved supersymmetry transformations.
The supersymmetry transformations for type II theories contain two 10-dimensional
Majorana-Weyl spinor parameters $\epsilon_{L}$ and $\epsilon_R$. Here, $L$ and $R$
indicate whether they originate from left- or right-moving
supersymmetry generators on the string world-sheet. In type IIA these spinors have
opposite chirality while in type IIB they have the same chirality.
In type II supergravity the supersymmetry transformations for the gravitino and dilatino read respectively:
\be
\begin{split}
\delta \psi_{L/R\,\mu} = & \left( \nabla_{\mu} \pm \frac{1}{4} \sla \! H_{\mu}\right) \epsilon_{L/R} = \nabla_{\pm\, \mu} \epsilon_{L/R}, \\
\delta \lambda_{L/R} = & \left(\sla \partial \Phi \pm \frac{1}{2}\sla \! H\right) \epsilon_{L/R},
\label{susy}
\end{split}
\ee
where $L$ gets the plus sign and $R$ the minus sign, $\nabla$ is the covariant derivative containing
the Levi-Civita connection, $\Phi$ is the dilaton, $H$ the NS-NS 3-form and all R-R forms were put to zero.
We consider geometries with both left- and right-moving preserved supersymmetries
generated by $\epsilon_L$ and $\epsilon_R$ respectively. The vanishing of the gluino supersymmetry transformation
on the brane will then relate $\epsilon_L$ and $\epsilon_R$.

If we introduce the sum of forms $\phi_0 \in \wedge^\bullet T_M^*$,
\be
\phi_0 = \sum_{l=0}^{10} \frac{1}{l!} \bar{\epsilon}_R \gamma_{\mu_1 \ldots \mu_l} \epsilon_L \, dx^{\mu_1} \wedge dx^{\mu_l},
\ee
we find using both $\delta \psi_{L/R\,\mu}=0$ and $\delta \lambda_{L/R}=0$ for $\epsilon_L$ and $\epsilon_R$ (see also \cite{jeschek,grana2})
\be
d \phi_0 - H \wedge \phi_0 - d \Phi \wedge \phi_0 =0.
\label{derphi0}
\ee
Therefore we should take our candidate generalized calibration to be
\be
\phi=e^{-\Phi} \phi_0 e^{-b},
\ee
so that it is closed. Furthermore we take $\pol(\cF)=e^{\cF}$ so that in the pull-back to the world-volume
$P(b)$ and $\cF$ indeed appear in the invariant combination $F=\cF-P(b)$.

To proceed we like to consider supersymmetric cycles in Euclidean geometry
so we split our space-time manifold as $\mathbb{R}^{1,9-d} \times M$ with Minkowski
metric on $\mathbb{R}^{1,9-d}$, Euclidean metric $g$ on the $d$-dimensional internal manifold, $H$ only non-vanishing on $M$
and everything independent of the coordinates in $\mathbb{R}^{1,9-d}$. We can then restrict ourselves to studying the Euclidean
geometry of the internal manifold $M$. The 10-dimensional Majorana-Weyl spinors $\epsilon_{L/R}$ decompose into spinors of $\mathbb{R}^{1,9-d}$
and spinors in the internal manifold. For instance in the case $d=2n$ we find
\be
\epsilon_{L/R}= \zeta_{L/R} \otimes \eta_{L/R} + \zeta^c_{L/R} \otimes \eta^c_{L/R}
\label{decomposition}
\ee
for any $(1,9-2n)$-dimensional Weyl-spinors $\zeta_{L/R}$, with $\zeta^c_{L/R}$ their Majorana
conjugates. We also have $\eta^c_{L/R}=C(\eta_{L/R})^*$ such that the $\epsilon_{L/R}$
are Majorana in 10 dimensions. Note that when $n$ is odd $\eta$
and $\eta^c$ have different chirality while when $n$ is even they have the same chirality.
Plugging \eqref{decomposition} into \eqref{susy} we find supersymmetry variations of
exactly the same form but now for the $\eta_{L/R}$ and $\eta^c_{L/R}$. If $\eta_{L/R}$ and $\eta^c_{L/R}$
generate independently preserved supersymmetries\footnote{This means there is no relation needed between $\eta_L$
and $\eta^c_L$ nor between $\eta_R$ and $\eta^c_R$ as would be the case in e.g.\ $Spin(7)_L \times Spin(7)_R$-structure.}
we can define
\be
\phi_{0,i_1\ldots i_l} = \sum_{l=0}^{d} \eta_R\dagg \gamma_{i_1 \ldots i_l} \eta_L, \quad \text{or} \quad \phi_{0,i_1\ldots i_l} = \sum_{l=0}^{d} \eta^c_R\dagg \gamma_{i_1 \ldots i_l} \eta_L
\label{calform}
\ee
and find that both also obey \eqref{derphi0}.

The case just presented, in which we have two preserved supersymmetries on the internal manifold on the left-moving side,
generated by $\eta_{L}$ and $\eta^c_{L}$ and two preserved supersymmetries on the right-moving
side generated by $\eta_R$ and $\eta^c_R$, will be the most studied in this paper. Normalizing the spinors such that
$\eta_{L/R}^\dagger \eta_{L/R}=1$, $\eta^c_{L/R}\dagg \eta^c_{L/R}=1$ we can define
\begin{subequations}
\begin{align}
\label{jdef}
& \omega_{L,ij} =-i \eta_L^\dagger \gamma_{ij} \eta_L, \qquad \omega_{R,ij} =-i \eta_R^\dagger \gamma_{ij} \eta_R, \\
\label{omegadef}
& \Omega_{L,i_1\ldots i_n} = \eta_L^\dagger \gamma_{i_1\ldots i_n}\eta^c_L, \qquad \Omega_{R,i_1\ldots i_n} = \eta_R^\dagger \gamma_{i_1\ldots i_n}\eta^c_R.
\end{align}
\end{subequations}
From this we can construct two almost complex structures $J_{L/R}=g^{-1}\omega_{L/R}$, $J_{L/R}^2=-\mathbf{1}$. It is possible to
show from the dilatino equation in \eqref{susy} that the Nijenhuistensors vanish \cite{strominger} so that $J_{L/R}$ are integrable.
Note that $\eta_{L/R}$ and $\eta^c_{L/R}$ are the empty and completely filled state of eqs.~\eqref{empty} and \eqref{filled} for
$J_{L/R}$ respectively.

From the vanishing of the gravitino transformations we find furthermore
\begin{subequations}
\begin{align}
\nabla^{\pm}_i \omega_{jk} & = \nabla^{\pm}_i \omega_{jk} \mp H_{i}{}^l{}_{[j}\omega_{|l|k]}=0, \label{nablaJ} \\
\nabla^{\pm}_i \Omega_{j_1\ldots j_n} & = \nabla^{\pm}_i \Omega_{j_1\ldots j_n} \mp \frac{n}{2} H_{i}{}^l{}_{[j_1}\Omega_{|l|j_2\ldots j_n]}=0 \label{nablaOm},
\end{align}
\end{subequations}
i.e. the left- and right-moving tensors are covariantly constant with respect to the Bismut connections $\nabla^+=\nabla + \frac{1}{2}H$
and $\nabla^-=\nabla - \frac{1}{2}H$ respectively. From the integrability of the complex structures $J_{L/R}$, their compatibility with the
metric $gJ+J^Tg=0$ and \eqref{nablaJ} follows that we have in fact bihermitian geometry $(g,J_L,J_R,H)$ \cite{bihermitian}. We
will use its connection to the generalized K\"ahler structure of \cite{hitchin,gualtieri} later in the paper. We have $U(3)_L \times U(3)_R$
structure which is further reduced to $SU(3)_L \times SU(3)_R$ structure by the existence of $\Omega_{L/R}$ satisfying \eqref{nablaOm}.
Since eqs.~\eqref{nablaJ} and \eqref{nablaOm} contain the Bismut connection instead of the Levi-Civita connection it does not follow that we
have special holonomy. Only when $H=0$ there is $SU(n)$ holonomy and $M$ is a Calabi-Yau manifold.

\subsection{The bound and the supersymmetry variation of the gluino}

In this subsection we establish the bound \eqref{bound} for our candidate generalized calibration $\phi$ and show that
the bound is saturated if and only if the gluino variation vanishes. In that case we say that the D-brane
$(N,\cF)$ wraps a supersymmetric cycle.

Let us define the following $\gamma$-matrix structures
\be
\begin{split}
\rho(F) & = \sum_l \frac{1}{2^l l!(p-2l)!} F_{a_1a_2} \ldots F_{a_{2l-1}a_{2l}} \gamma_{a_{2l+1}\ldots a_{p}} \epsilon^{a_1 \ldots a_p},\\
\Gamma(F) & = \frac{1}{\sqrt{\det(P(g)+F)}} \rho(F),
\label{gammaF}
\end{split}
\ee
with as before $F=\cF - P(b)$.
Using the methods of \cite{aganagic,cederwall,bergshoeff} we can show that
\be
\rho(F)^{\dagger}\rho(F) = (\rho_E(F) + \rho_O(F))(\rho_E(F)-\rho_O(F))= \det (P(g)+F),
\label{square}
\ee
where $(\rho_E(F))^\dagger=\rho_E(F)$ and $\rho_O(F)^\dagger=-\rho_O(F)$ the hermitian and anti-hermitian
part of $\rho(F)$. Alternatively we have $\Gamma(F)^\dagger \Gamma(F)=1$.

These matrices are closely related to the $\Gamma$-matrix defined in \cite{aganagic,cederwall,bergshoeff}. That matrix
plays a crucial role in the definition of the $\kappa$-symmetry and supersymmetry transformations for D$p$-branes.
In fact, for the D$p$-branes we consider in this paper, i.e.\ the ones extended solely in the internal manifold with only magnetic fields turned on, we have
\be
\Gamma_{\text{IIA}}=-(\gamma_{11} \gamma_0 \Gamma_E+\gamma_0 \Gamma_O), \qquad
\Gamma_{\text{IIB}}=-(\gamma_0 \Gamma_E+\tau_3 \gamma_0 \Gamma_O)\tau_1.
\ee
These $\gamma$-matrix structures convert left-moving spinors into right-moving spinors. Indeed, in the IIA case $\Gamma_{\text{IIA}}$ contains an
odd number of $\gamma$-matrices so that it changes the chirality while in the IIB case $\tau_1$ takes care of the switch.
They also satisfy $\Gamma^\dagger=\Gamma$ and $\Gamma^2=1$. The gluino supersymmetry transformation in a certain
$\kappa$-gauge consists of a supersymmetry transformation and a compensating $\kappa$-transformation. In \cite{bergshoeff2}
it is shown that the preserved supersymmetries must satisfy
\be
(1-\Gamma)\epsilon=0,
\ee
with $\epsilon=(\epsilon_L,\epsilon_R)$.
For the part of the spinors on the internal manifold this translates into
\be
\Gamma(F)\eta_L = e^{i \gamma} \eta_R,
\label{gluinosusy}
\ee
with $e^{i \gamma}$ a constant phase.

Using eq.~\eqref{square} we can, following \cite{thesis}, link the gluino supersymmetry condition \eqref{gluinosusy}
to the bound \eqref{bound}. Indeed
\be
\begin{split}
\det(P(g)+F) & = \eta_L^\dagger \det(P(g)+F) \eta_L
             = \sum_{\eta'} \left(\eta_L^\dagger \rho^\dagger(F) \eta' \right) \, \left( \eta'{}^\dagger \rho(F) \eta_L\right)
             = \sum_{\eta'} \left|\eta'{}^\dagger \rho(F) \eta_L\right|^2 \\
             & \ge \left|\eta_R^\dagger \rho(F) \eta_L\right|^2 \ge \left(\Re \left( e^{-i \gamma} \eta_R^\dagger \rho(F) \eta_L \right)\right)^2,
\end{split}
\ee
with $\gamma$ the same constant as before.
In the second line we have introduced an orthonormal complete set of spinors $\sum_{\eta'} \eta' \, \eta'{}^\dagger=1$.
In the end we find the bound
\be
e^{-\Phi} \sqrt{\det(P(g)+F)} \ge \Re \left( e^{-i \gamma} e^{-\Phi} P(\phi_0) e^{F}\right)_\epsilon,
\ee
where $\phi_0$ given by \eqref{calform}. Moreover, from \eqref{derphi0} we know that
$d \left(\Re \left( e^{-i \gamma} e^{-\Phi} \phi_0 e^{-b} \right)\right)=0$ so that we
have indeed constructed a generalized calibration.

The bound is saturated if and only if
\begin{subequations}
\begin{align}
\label{cond1}
& \eta'{}^\dagger \rho(F) \eta_L =0, \quad \forall \eta' \neq \eta_R,\\
\label{cond2}
& \sqrt{\det(P(g)+F)} = e^{-i \gamma} \eta_R^\dagger \rho(F) \eta_L.
\end{align}
\end{subequations}
These two conditions are completely equivalent to \eqref{gluinosusy}. It follows that
this type of generalized calibrated D-branes is supersymmetric and vice-versa every supersymmetric
D-brane is a generalized calibrated D-brane of this type.

Another related viewpoint on supersymmetry vs.\ calibrations made of bilinears of spinors is based on
central charges in the supersymmetry algebra \cite{wittenolive,olive,townsend1,townsend2}.
The calibration bound is then the well-known BPS bound and when there is unbroken supersymmetry the Hamiltonian
is equal to the central charge. This approach is heavily used in \cite{gencal3}. We defer working out the details
for the calibrations at hand to further work.

\section{World-sheet approach}
\label{worldsheet}

In this section we consider the conditions for unbroken target space supersymmetry again, but now
in the string world-sheet approach. We will find exactly the same conditions \eqref{cond1} and \eqref{cond2}.
The special case of $H=0$ was already studied from this viewpoint in \cite{kapustinstable}.

Here we study an $N=(1,1)$ non-linear sigma-model with bulk metric $g$ and torsion $H=db$.
As has been found in \cite{ooguri} and later studied in great detail
in \cite{albertsson,susyboundary}, if we introduce
a D-brane $(N,\cF)$, the gluing conditions read
\be
\psi_R^i= R^i{}_j \psi_L^j,
\label{gluing}
\ee
with
\be
R = P_+ \frac{g - F}{g + F} P_+ - P_-.
\ee
Here the submanifold $N$ on which the D-brane wraps is determined by the integrable product structure
$r=P_+ - P_-$, $r^2=\mathbf{1}$, which is compatible with the metric, i.e.\ $r^T g r =g$.
$P_+$, satisfying $P_+^2=P_+$, projects on vectors tangential to the D-brane, while $P_-$, satisfying $P_-^2=P_-$, projects on
vectors normal to the D-brane. We also have
\be
F = \cF_M - P_+ b P_+,
\label{extF}
\ee
where $\cF_M$ is a smooth extension of $\cF$ to $M$, i.e.\ $P(\cF_M)=\cF$, satisfying $P_- \cF_M =\cF_M P_-=0$.

We can promote the $N=(1,1)$ supersymmetry to an $N=(2,2)$ supersymmetry if and only if the target space manifold $M$
admits a bihermitian geometry $(g,J_L,J_R,H)$ \cite{bihermitian}. The $U(1)$ R-currents of the $N=(2,2)$ geometry
read
\be
j_L = \omega_{L,ij} \psi_L^i \psi_L^j, \qquad j_R = \omega_{R,ij} \psi_R^i \psi_R^j.
\ee
If the D-brane is to preserve $N=2$ supersymmetry we must have $j_L=j_R$ on the boundary. Using eq.~\eqref{gluing}
we find that we must have
\be
R^T \omega_{R} R = \omega_L,
\label{omegacond}
\ee
or alternatively
\be
R^{-1} J_R R = J_L.
\label{Jcond}
\ee
Plugging \eqref{jdef} into \eqref{omegacond} the condition becomes
\be
\eta_R^\dagger R^i{}_k R^j{}_l \gamma_{ij} \eta_R = \eta_L^\dagger \gamma_{kl} \eta_L.
\label{omegacond2}
\ee

Before proceeding, we will first show that the $\Gamma(F)$ defined in \eqref{gammaF} is in fact the spinor
representation of $R$. We rewrite $\Gamma(F)$ as:
\be
\Gamma(F)=\frac{\sqrt{\det(P(g))}}{\sqrt{\det(P(g)+F)}} \, \mathrm{se} (- \sla \! F) \Gamma_N,
\label{gammaF2}
\ee
with ``se'' the skew-exponential function (the usual exponential function but with $\gamma$-matrices completely
symmetrized at every order):
\be
\mathrm{se}(\sla \! F) = \sum_{l=0}^{\lfloor p/2 \rfloor} \frac{1}{2^l l!} \gamma^{a_1 \ldots a_{2l}} F_{a_1a_2}\ldots F_{a_{2l-1}a_{2l}} = \sla \! e^F,
\ee
and
\be
\Gamma_N = \frac{1}{p! \sqrt{\det(P(g))}} \epsilon^{a_1 \ldots a_p}\gamma_{a_1 \ldots a_p}.
\ee
In \cite{bergshoeff2} it is shown that
\be
\frac{\sqrt{\det(P(g))}}{\sqrt{\det(P(g)+F)}} \mathrm{se} \left(- \sla \! F\right) = \exp\left[ -\frac{1}{4} \phi^{ab} \gamma_{ab}\right],
\ee
with $\phi=2 \arctan F = \ln \frac{P(g)+F}{P(g)-F}$. This is a rotation with angle matrix $\phi$ in the spinor representation.
On the other hand, $\Gamma_N (\gamma_{d+1})^{p+1}$ is the spinor representation of a reflection in the directions normal to the D-brane.
Taking both together we find that if we define the spinor representation $U_R$ as
\be
R^i{}_j \gamma_i = U_R^\dagger \gamma_j U_R,
\label{spinorrep}
\ee
then
\be
U_R^\dagger = U_R^{-1} = \Gamma(F) \left(\gamma_{d+1}\right)^{p+1}.
\label{Rspinorrep}
\ee

Picking up where we left off at eq.~\eqref{omegacond2} and plugging in \eqref{spinorrep}
we find
\be
\left(U_R \eta_R \right)^\dagger \gamma_{kl} \left(U_R \eta_R\right) = \eta_L^\dagger \gamma_{kl} \eta_L
\label{omegacond3}
\ee
Using eq.~\eqref{Rspinorrep} and rephrasing \eqref{cond1} as
\be
\Gamma(F) \eta_L = e^{i \beta(\sigma)} \eta_R,
\label{cond1rephrase}
\ee
we see that condition \eqref{cond1} implies \eqref{omegacond3} and thus \eqref{omegacond}. Since both $\eta_L$ and $\eta_R$ are
normalized, the proportionality factor $e^{i\beta(\sigma)}$ can indeed only be a phase. At this point, it may still vary over the D-brane though.
We emphasize this by indicating that it can be a function of the D-brane world-volume coordinates $\sigma$.
It is condition \eqref{cond2} that will fix it to a constant phase $\beta(\sigma)=\gamma$.

Conversely, from \eqref{Jcond} follows that for every vector $v$ that is a $(+i)$-eigenvalue
of $J_L$, $w=Rv$ is a $(+i)$-eigenvalue of $J_R$, and vice-versa. Now $\eta_R$ is the spinor that is annihilated by
all $w^i\gamma_i$ with $w$ a $(+i)$-eigenvalue of $J_R$. Using eq.~\eqref{spinorrep} it follows that $U_R \eta_R$ is
annihilated by all $v^i \gamma_i$ with $v$ a $(+i)$-eigenvalues of $J_L$. This implies $U_R \eta_R \propto \eta_L$ or equivalently
\eqref{cond1rephrase}.

Summarizing we find:
\be
\begin{split}
R^{-1} J_R R & = J_L \Leftrightarrow \eta'{}^\dagger \rho(F) \eta_L =0, \quad \forall \eta' \neq \eta_R,\\
R^{-1} J_R R & = -J_L \Leftrightarrow \eta'{}^\dagger \rho(F) \eta_L =0, \quad \forall \eta' \neq \eta^c_R,
\label{cond1rephrase2}
\end{split}
\ee
The second statement can be proven analogously or just by noting that changing $J \rightarrow -J$
will indeed send $\eta \rightarrow \eta^c$. For later use we also note that
\be
\rho(F) \eta_L \propto \eta_R \Leftrightarrow R^{-1} J_R R = J_L \Leftrightarrow R^{-1} (-J_R) R = -J_L \Leftrightarrow \rho(F) \eta^c_L \propto \eta^c_R.
\label{gammaupspinor}
\ee

From the point of view of topological string theory a boundary condition that preserves $N=2$ supersymmetry is
a {\em topological D-brane}. This is however not enough to have unbroken supersymmetry for the D-branes in target space. So, also from
the world-sheet analysis we find a second condition, which we will derive now. The target space supersymmetry is generated
by the spectral flow operators
\be
S_{L/R}= \frac{1}{n!} \Omega_{L/R,i_1\ldots i_n} \psi_{L/R}^{i_1} \ldots \psi_{L/R}^{i_n}.
\ee
On the boundary we require matching of the spectral flow operators
\be
S_R = e^{i \alpha} S_L.
\label{specmatch}
\ee
We would like to show that this matching condition  is equivalent to \eqref{cond2} and find the precise relation between the phases $e^{i\alpha}$ and
$e^{i\gamma}$. In order to do that, we introduce a charge conjugation matrix $C$ (see \eqref{conjugation} for the defining property) such that
\be
\eta_L^c=C \eta_L^*.
\label{leftup}
\ee
Plugging in \eqref{omegadef} and using the same trick as above we rewrite the matching condition \eqref{specmatch} as
\be
\left(U_R \eta_R \right)^\dagger \gamma_{i_1\ldots i_n} U_R \eta^c_R = e^{i \alpha} \eta_L^\dagger \gamma_{i_1\ldots i_n} \eta^c_L
\label{Omegacond2}
\ee
Suppose the condition \eqref{cond1rephrase} for having a topological brane is already satisfied. If we take the chirality of $\eta_L$
to be positive we can rewrite it as
\be
U_R \eta_R = e^{-i\beta(\sigma)} \eta_L.
\ee
Let us now calculate
\be
\Gamma(F) \eta_L^c = \Gamma(F) C \eta_L^*=C \Gamma(F)^* \eta_L^* = e^{-i\beta(\sigma)} C \eta_R^*,
\ee
where we used \eqref{leftup}, \eqref{conjugation} and \eqref{cond1rephrase}. We already know from \eqref{gammaupspinor}
that this result should be proportional to $\eta^c_R$. However, there is some phase arbitrariness in the
definition of $\eta_R$ and thus also in the definition of $\gamma$ in \eqref{cond2}. In order to find a definite relation
between $\alpha$ and $\gamma$ we fix it by choosing
\be
\eta_R^c=C \eta_R^*,
\label{rightup}
\ee
with the {\em same} charge conjugation matrix as used for the left-movers,
and find
\be
\Gamma(F) \eta_L^c = e^{-i\beta(\sigma)} \eta^c_R,
\ee

From this follows
\be
U_R \eta_R^c = (-1)^n e^{i \beta(\sigma)} \eta_L^c.
\ee
Plugging this into \eqref{Omegacond2} we find that $\beta(\sigma)=\gamma$ should be constant and
find the relation
\be
e^{i \alpha} = (-1)^n e^{2 i \gamma}.
\ee

Concluding, the world-sheet supersymmetry conditions \eqref{Jcond} and \eqref{specmatch} are exactly equivalent
to the generalized calibration conditions \eqref{cond1} and \eqref{cond2} which are in turn equivalent to the
target space gluino supersymmetry condition \eqref{gluinosusy}. The condition \eqref{specmatch} is called the {\em stability}
condition.

\section{Special case: Calabi-Yau manifold}
\label{CY}

In this section we specialize to the case $H=0$, but non-vanishing $F$, and present the two canonical examples studied
before in \cite{minasian}
from the target space perspective and in \cite{kapustinstable} from the world-sheet perspective. In this case we find:
\be
\nabla J_{L/R} = d\omega = \nabla \Omega_{L/R} = d \Omega_{L/R} = \nabla \eta_{L/R}=0,
\ee
where $\nabla$ is the covariant derivative containing the Levi-Civita connection. So we have $SU(3)$-holonomy and $M$ is
a Calabi-Yau manifold. If we do not introduce any extra special holonomy we find that
either $J_R=J_L$ or $J_R=-J_L$. Since mirror symmetry reverses the sign of $J_R$ it will exchange
these two cases.

\subsection{B-branes: the complex case}

In the first case of $J_R=J_L$, we find
\be
R^{-1} J R = J,
\ee
from which follows $r J r = J$, which implies $N$ is a complex submanifold with complex structure $P_+ J P_+$, and $F$ is of type
$(1,1)$ on $N$. We must have $p=2k$ even.
We take $\eta_L=\eta_R$ and find for the calibration condition \eqref{cond2}:
\be
\sqrt{\det(P(g)+F)} = e^{-i \gamma} \frac{1}{k!} (iP(\omega)+F)^k|_\epsilon=i^{k-n} e^{-i \alpha/2} \frac{1}{k!}(P(\omega)-iF)^k|_\epsilon.
\ee
For small $F$ this reduces to the Donaldson-Uhlenbeck-Yau condition $g^{\alpha\bar{\beta}} F_{\alpha\bar{\beta}}=0$.
The case of $p=d$ was extensively studied in \cite{thesis} (see also references therein). In there, the existence of
a calibration bound like this was used as a constraint to calculate derivative and non-abelian
corrections to the D-brane effective action.

\subsection{A-branes: the coisotropic case}

In the second case, we find
\be
R^{-1} J R = -J.
\ee
As was first shown in \cite{kapustincoisotropic} this
leads to the following three properties (see also section 7.2 of \cite{gualtieri}):
\begin{enumerate}
\item Let us define $\Ann T_N=\{\xi \in T^*_M\, : \, X^i \xi_i =0, \, \forall X \in T_N\}$. Then we
have $\omega^{-1}(\xi) \in T_N, \forall \xi \in \Ann T_N$. This means the submanifold is {\em coisotropic}.
It also implies that the symplectic orthogonal bundle $T_N^\perp=\{Y \in T_M \, : \, \omega_{ij}X^i Y^j =0, \, \forall X \in T_N\}$
lies within $T_N$: $T_N^\perp \subseteq T_N$. Because $\omega$ is non-degenerate, the dimension of $T_N^\perp$
is the codimension of $T_N$.
\item $F Y=0, \, \forall Y \in T_N^\perp$ i.e. $F$ descends to $T_N/T_N^\perp$.
\item $\left(\omega|_N\right)^{-1}F$ is an almost complex structure on $T_N/T_N^\perp$. In fact, in \cite{kapustincoisotropic}
it was shown that this complex structure is integrable.
\end{enumerate}
It can be shown that the complex dimension of $T_N/T_N^\perp$ should be even. This implies that $p-n=2k$ is even.
We take now
\be
\eta_R = (-1)^{\frac{n(n-1)}{2}} \eta^c_L.
\ee
The phase factor comes about because we want to stick to our convention \eqref{rightup}.
In calculating it we used \eqref{cinvariantsign}.
For the calibration condition \eqref{cond2} we find
\be
\sqrt{\det(P(g)+F)} = e^{i \gamma} \left(P(\Omega) \wedge e^{F}\right)_{[p],\epsilon}.
\ee

\section{Generalized Calabi-Yau geometry}
\label{gencomp}

In \cite{gualtieri} it was shown that bihermitian geometry is equivalent to generalized K\"ahler structure.
In the latter context, yet another notion of generalized calibrations was introduced.
We will show however that these calibrations exactly coincide
with the generalized calibrations studied here. First we start with a lightning review of generalized complex
geometry and connect these concepts to the ideas of the previous sections as they are introduced.
The reader is advised however to also consult \cite{hitchin} and \cite{gualtieri}.

\subsection{Generalized complex geometry}

In generalized (complex) geometry the usual statements about integrable subbundles of the tangent bundle $T_M$
are replaced by similar statements about subbundles of $T_M \oplus T^*_M$. On this space there exists a natural metric
defined by $(U,V)=(X+\xi,Y+\eta)=\frac{1}{2}\left(i_X \eta + i_Y \xi\right)=\frac{1}{2}\left(X^i \eta_i + Y^i \xi_i\right)$
with $X,Y \in T_M$ and $\xi,\eta \in T^*_M$.
We will denote this metric, which has $(d,d)$-signature, by $I$.
A subbundle $L \subset T_M \oplus T^*_M$ is {\em isotropic} if $(U,V)=0$ for every $U,V \in L$. If it has the maximal dimension $d$, it is called
{\em maximally isotropic}. A subbundle $L$ is {\em integrable} if the
{\em Courant bracket},
\be
[U,V]=[X+\xi,Y+\eta]=[X,Y]_L+{\cal L}_X \eta - {\cal L}_Y \xi - \frac{1}{2} d (i_X \eta - i_Y \xi),
\ee
is closed on $L$, i.e.\ if it is involutive. Here, $[.,.]_L$ is the Lie bracket on $T_M$ and ${\cal L}$ is the Lie-derivative.
The Courant bracket can be twisted by a closed 3-form $H$ as follows
\be
[U,V]_H=[X+\xi,Y+\eta]_H=[X,Y]_L+{\cal L}_X \eta - {\cal L}_Y \xi - \frac{1}{2} d (i_X \eta - i_Y \xi) + i_X i_Y H.
\ee
A subbundle $L$ that is involutive under the $H$-twisted Courant bracket is called $H$-integrable.
The only symmetries of the Lie bracket are diffeomorphisms. The Courant bracket however has an extra
symmetry which is called the $b$-transform\footnote{Note that our sign convention for the $b$-transform differs from
the one in the generalized complex structure literature. In this convention one can represent the $b$-transform
as the matrix $\left(\begin{array}{cc} 1 & 0 \\ b & 1\end{array}\right)$ working on $(X \; \xi)^T$.}:
\be
e^b(X+\xi)=X+\xi-i_X b = X + \xi + b X.
\ee
Under the $b$-transform the Courant bracket changes as
\be
[e^b U, e^b V]_H = e^b [U,V]_{H+db}.
\label{bcourant}
\ee
Therefore the $b$-transform is an automorphism if and only if $db=0$.
Taking $H'=H+db$ in eq.~\eqref{bcourant} we see that if $L$ is $H'$-integrable,
then $e^b L$ will be $(H'-db)$-integrable.

An element $U \in T_M \oplus T^*_M$ has a natural action on a sum of forms of different dimensions
(henceforth just called form), $\phi \in \wedge^\bullet T_M^*$ as follows:
\be
U \cdot \phi = (X,\xi)\cdot \phi = i_X \phi + \xi \wedge \phi.
\label{actionspinor}
\ee
In fact, this makes $T_M \oplus T^*_M$ a realization of the Clifford algebra $\mathrm{Cliff}(d,d)$ and the forms the spin
representation since $(X+\xi)^2 \cdot \phi = (i_X \xi) \phi = (X+\xi,X+\xi)\phi$. A spinor $\phi$ is called
{\em pure} if its {\em null space} $L_{\phi}=\{U \in T_M \oplus T^*_M \, : \, U \cdot \phi=0\}$ is maximally isotropic.
Every maximally isotropic subbundle $L$ is represented by a unique pure spinor line $U_L$ (i.e. a spinor
defined up to a proportionality factor). If $\phi$ is a pure spinor of $L$, then $e^b \phi$ will be a pure spinor
of $e^b L$. Pure spinors have a definite {\em parity}. They are positive if they consist solely of even forms
and negative if they consist of odd forms.

In \cite{hitchin} it was shown that $L$ is $H$-integrable if and only if for any spinor $\phi$
of the corresponding pure spinor line there exists a $U=(X,\xi)$ such that it satisfies
$d_H \phi=(d+H\wedge) \phi=i_X \phi + \xi \wedge \phi$. In many examples, it will be possible
to find a pure spinor such that simply $d_H \phi=0$. In that case we find indeed that $d_{H-db}(e^b L)=0$
such that $e^b L$ is $(H-db)$-integrable.

Let us now interpret eq.~\eqref{gluing} in terms of a maximally isotropic subbundle. The $(1,1)$-tensor
$r$ defines a distribution $E \subseteq T_M$ consisting of the vector fields $v$ satisfying $r v=v$.
If the distribution is involutive with respect to the Lie bracket, through a point $x$ we can define a
submanifold $N$ such that $E|_N=T_N$. We consider now a D-brane $(N,\cF)$ and define $F$ as in eq.~\eqref{extF}.
Following
\cite{zabzineboundary} we introduce
\be
\psi^{i} = \psi_L^i + \psi_R^i, \qquad \rho_{i} = g_{ij} \left( \psi_L^j - \psi_R^j\right).
\label{psirho}
\ee
The gluing condition \eqref{gluing} can then be rewritten as
\be
{\mathcal{R}} \Psi = \Psi,
\ee
with
\be
\mathcal{R}= e^F r e^{-F} = \left( \begin{array}{cc} \mathbf{1} & 0\\F & \mathbf{1}\end{array}\right)
\left( \begin{array}{cc} r & 0\\0 & -r^t\end{array}\right)
\left( \begin{array}{cc} \mathbf{1} & 0\\-F & \mathbf{1}\end{array}\right).
\label{calR}
\ee
and $\Psi= \left( \psi \; \rho \right)^T$. In fact, this means that $\Psi$ belongs
to the generalized tangent bundle $(N,F)$, which is defined as
\be
L(N,F)=\{X+\xi \in T_N \oplus T_M^* \, : \, \xi|_N = i_X F\}=e^F \left(T_N \oplus \Ann T_N\right).
\ee
This is a maximally isotropic subbundle. In \cite{gualtieri} it is shown that
the generalized tangent bundle is involutive with respect to the
$H$-twisted Courant bracket precisely if $E$ is involutive and $dF=-H|_N$.
The corresponding pure spinor is given by
\be
\tau_{(N,F)} = c \exp F \det \Ann T_N,
\label{subpurespinor}
\ee
where $\det \Ann T_N = \theta_1 \wedge \ldots \wedge \theta_{d-p}$ with $(\theta_1,\ldots,\theta_{d-p})$
a basis for $\Ann T_N$. $c \neq 0$ is an arbitrary constant.
We see that in fact $\rho(F)$ defined in eq.~\eqref{gammaF} and rewritten in the manner of eq.~\eqref{gammaF2}
reads
\be
\rho(F) = \sqrt{P(g)} \mathrm{se}(- \sla \! F) \Gamma_N = \sqrt{P(g)} \mathrm{se}(- \sla \! F) \Gamma_N^\perp \gamma_{1 \ldots d},
\label{rhoF2}
\ee
where we defined $\Gamma_N^\perp \gamma_{1 \ldots d} = \Gamma_N$. By going to a coordinate system where the tangent
directions to the submanifold are denoted by the first $p$ coordinates, one easily sees that $\Gamma_N^\perp$ will
be a product of $\gamma$-matrices in the normal space. Eqs.~\eqref{subpurespinor} and \eqref{rhoF2} show
that $\tau_{(N,F)}$, the pure spinor associated to the generalized tangent bundle defined by $\mathcal{R}$,
and $\rho(F)$, the spinor representation of the $R$ in \eqref{gluing} are closely related. To be precise:
\be
\rho(F)=\sqrt{P(g)} (-1)^{\frac{(d-p)(d-p-1)}{2}} \sum_l \frac{1}{l!} \left(\dual{\tau}_{(N,F)}\right)^{j_{1}\ldots j_{l}} \gamma_{j_1 \ldots j_l}.
\label{rhoF3}
\ee

An {\em almost generalized complex structure} is a map $\mathcal{J}: T_M \oplus T^*_M \rightarrow T_M \oplus T^*_M$ such that $\mathcal{J}^2=-\mathbf{1}$
and that $\mathcal{J}$ is compatible with the metric: $(\mathcal{J}U,\mathcal{J}V)=(U,V)$. Let $L$ and $\bar{L}$
denote the $(+i)$ and $(-i)$-eigenbundles of $\mathcal{J}$ respectively.  $L$ and $\bar{L}$ are maximally isotropic
subbundles. $\mathcal{J}$ is an $H$-twisted generalized complex structure if and only if $L$ is $H$-integrable (which implies that
$\bar{L}$ is also $H$-integrable). We denote the pure spinor associated to $L$ by $\phi_\mathcal{J}$.

An {\em $H$-twisted generalized K\"ahler structure} is a pair $(\mathcal{J}_1,\mathcal{J}_2)$ of commuting $H$-twisted
generalized complex structures such that $G= I F = -I \mathcal{J}_1 \mathcal{J}_2$ is a positive definite metric on $T_M \oplus T^*_M$.
Note that $F=- \mathcal{J}_1 \mathcal{J}_2$ satisfies $F^2=\mathbf{1}$. We will call the $(+1)$ and $(-1)$-eigenbundles
of $F$, $C_+$ and $C_-$ respectively. We can define the projections $p_{\pm} = \frac{1}{2}(1+F)$ on $C_{\pm}$ and
the projection $\pi$ on $T_M$ such that $\pi(X,\xi)=X$. It is easy to see that $\mathcal{J}_1=\mathcal{J}_2$ on $C_+$
and $\mathcal{J}_1=-\mathcal{J}_2$ on $C_-$. By projection from $C_{\pm}$, $\mathcal{J}_1$ induces two almost complex
structures on $M$, which we denote $J_{L/R}$. More concretely, they are defined such that
\be
\begin{split}
\mathcal{J}_1 & = \pi|_{C_+}^{-1} J_L \pi p_+ + \pi|_{C_-}^{-1} J_R \pi p_-, \\
\mathcal{J}_2 & = \pi|_{C_+}^{-1} J_L \pi p_+ - \pi|_{C_-}^{-1} J_R \pi p_-.
\end{split}
\label{JLJR}
\ee
We see that since mirror symmetry sends $J_R \rightarrow -J_R$ it interchanges $\mathcal{J}_1$ and $\mathcal{J}_2$.
In coordinates the metric $G=I F$ has the form
\be
F= e^b g e^{-b} = \left( \begin{array}{cc} \mathbf{1} & 0\\b & \mathbf{1}\end{array}\right)
\left( \begin{array}{cc} 0 & g^{-1} \\g & 0\end{array}\right)
\left( \begin{array}{cc} \mathbf{1} & 0\\-b & \mathbf{1}\end{array}\right),
\label{genmetric}
\ee
and is thus determined by the pair $(g,b)$. Furthermore
\be
\mathcal{J}_{1/2} = \frac{1}{2} \, e^b \left( \begin{array}{cc} J_L \pm J_R & -(\omega_{L}^{-1}\mp \omega_R^{-1}) \\
\omega_{L} \mp \omega_R & -(J_L^T \pm J_R^T) \end{array}\right)
 e^{-b},
\label{J1J2}
\ee
with as before $\omega_{L/R}=g J_{L/R}$. The two pure spinors $\mathcal{J}_1$ and $\mathcal{J}_2$ are of the same
parity if $n=d/2$ even and of opposite parity if $n$ odd.

In \cite{gualtieri} it is shown that the generalized K\"ahler geometry $(\mathcal{J}_1,\mathcal{J}_2)$ is completely
equivalent to the bihermitian geometry $(g,J_L,J_R,H)$ namely $\mathcal{J}_1$ and $\mathcal{J}_2$ defined in \eqref{J1J2}
are $\tilde{H}$-integrable with $\tilde{H}=H-db$ if and only if the corresponding $J_L$ and $J_R$ from \eqref{JLJR}
satisfy
\be
(\nabla \pm \frac{1}{2} H) J_{L/R} = 0,
\ee
and $H$ is of type $(2,1)+(1,2)$ with respect to both $J_{L/R}$ (the latter condition is implied by $J_{L/R}$ integrable).
Note that to each bihermitian structure correspond many generalized K\"ahler structures which differ from each other by
a $b$-transform. In what follows we apply a $b$-transform such that $b=0$ in \eqref{genmetric} and \eqref{J1J2}. This means
that we will put all $b$-dependence into $\mathcal{R}$ defined in \eqref{calR}.

The canonical example is a usual K\"ahler structure $(g,J,\omega)$ with $\omega=g J$.
The generalized complex structures and metric are
\be
\mathcal{J}_1 = \left( \begin{array}{cc} J & 0 \\
0 & -J^T \end{array}\right), \qquad \mathcal{J}_2 = \left( \begin{array}{cc} 0 & -\omega^{-1} \\
\omega & 0 \end{array}\right), \qquad G= - I \mathcal{J}_1 \mathcal{J}_2 = \left( \begin{array}{cc} g & 0 \\
0 & g^{-1} \end{array}\right).
\ee
$\mathcal{J}_1$ is Courant integrable if and only if $J$ is Lie integrable and $\mathcal{J}_2$ is integrable
if and only if $d \omega = 0$ so that we indeed find a K\"ahler structure. The corresponding pure spinors are
$\bar{\Omega}$ and $e^{-i \omega}$. From eq.~\eqref{J1J2} we find $J_L = J_R$.
Note that putting $J_L = - J_R$ corresponds to switching $\mathcal{J}_1 \leftrightarrow \mathcal{J}_2$.
These are the two special cases under study in section \ref{CY}.

With the choice $b=0$, vectors of $C_{\pm}$ have the form $(v^a,\pm g_{ab}v^b)$ respectively. From eq.~\eqref{psirho}
we see that $\Psi = (\psi, \rho)=(\psi_L, g \psi_L) + (\psi_R, -g \psi_R)$ so that the part with $\psi_L$ belongs
to $C_+$ and the part with $\psi_R$ to $C_-$.  In \cite{zabzineboundary} it is shown that
\be
R^{-1} J_R R = \pm J_L \Leftrightarrow \mathcal{J}_{1/2} = \mathcal{R}^{-1} \mathcal{J}_{1/2} \mathcal{R}.
\ee
So using eq.~\eqref{cond1rephrase2} we see that the first condition for the saturation of the calibration bound \eqref{cond1}
means that the generalized tangent bundle $L(N,F)$ of the D-brane should be invariant under $\mathcal{J}_{1/2}$
or $\mathcal{J}_2$. This implies the generalized submanifold should be in fact a generalized complex submanifold. As studied in
section \ref{CY} in the case of a Calabi-Yau manifold this leads to complex and coisotropic D-branes respectively.

We will study the second condition in the next subsection, but first we need to introduce the pure spinors of $\mathcal{J}_{1/2}$.
As in \cite{grana1,grana2}
it will be convenient to explore the connection between forms on $M$ and $\mathrm{Cliff}(d)$ bispinors:
\be
C = \sum_l \frac{1}{l!} C^{[l]}_{i_1 \ldots i_l} dx^{i_1}\wedge \ldots \wedge dx^{i_l}
\longleftrightarrow \sla \! C = \sum_l \frac{1}{l!} C^{[l]}_{i_1\ldots i_l} \gamma^{i_1 \ldots i_l}.
\ee
Because of \eqref{actionspinor} we have a relation between $\mathrm{Cliff}(d,d)$ spinors and sums of forms which
are thus in turn related to $\mathrm{Cliff}(d)$ bispinors. We can act on the bispinor with $\gamma$-matrices from
the left (denoted by $\overrightarrow{\gamma_i}$) and $\gamma$-matrices from the right (denoted by $\overleftarrow{\gamma_i}$),
which identifying $C$ and $\sla \! C$ has the effect:
\be
\begin{split}
X^j & \overrightarrow{\gamma_j} \longleftrightarrow X^j i_j + g_{ij} X^j dx^i \wedge = (X,g X)\cdot, \\
X^j & \overleftarrow{\gamma_j} \longleftrightarrow (-1)^{P+1} \left(X^j i_j - g_{ij} X^j dx^i \wedge \right)= (-1)^{P+1}(X,-g X)\cdot,
\end{split}
\ee
with $(-1)^P$ the parity of the pure spinor. So we see that $X^j \overrightarrow{\gamma_j}$ reproduces the action
of elements of $C_+$ and $X^j \overleftarrow{\gamma_j}$ of elements of $C_-$. It follows immediately
that $\eta_L \, \eta_R^\dagger$ is annihilated by $(+i)$-eigenvalues of $J_L$ on the $C_+$-side and $(-i)$-eigenvalues
of $J_R$ on the $C_-$-side. We see from eq.~\eqref{JLJR} that it must corresponds to $\mathcal{J}_2$. Analogously
$\eta_L \left(\eta^c_R\right)^\dagger$ corresponds to $\mathcal{J}_1$. Fierzing one finds
\be
\sla \! \phi_{\mathcal{J}_2}= \eta_L \, \eta_R^\dagger = \frac{1}{\mathrm{dim}\, S} \sum_l \frac{(-1)^{\frac{l(l-1)}{2}}}{l!} \left(\eta_R^\dagger \gamma_{i_1\ldots i_l} \eta_L\right) \gamma^{i_1\ldots i_l},
\label{phiJ}
\ee
with $\mathrm{dim}\, S$ the dimension of the spinor representation. We find the analogous expression for $\mathcal{J}_1$ by replacing
$\eta_R \rightarrow \eta^c_R$.
From eq.~\eqref{derphi0} it follows that
\be
d \left(e^{-\Phi} \phi_{\mathcal{J}_{1/2}}\right) + H \wedge e^{-\Phi} \phi_{\mathcal{J}_{1/2}}=0,
\label{CYhitchin}
\ee
which means that both generalized complex structures are not only $H$-integrable but in fact $H$-twisted generalized Calabi-Yau
structures according to Hitchin's definition \cite{hitchin}. Note that Hitchin's definition does not require a generalized
K\"ahler structure, just one generalized complex structure.

However, we can also show that we have a generalized Calabi-Yau structure in the sense of Gualtieri \cite{gualtieri},
which is rather a generalized K\"ahler structure so that both pure spinors $e^{-\Phi} \phi_{\mathcal{J}_{1/2}}$ satisfy
eq.~\eqref{CYhitchin} and
their lengths are related by a constant $c \in \mathbb{R}$:
\be
(\phi_{\mathcal{J}_{1}},\bar{\phi}_{\mathcal{J}_{1}})=c(\phi_{\mathcal{J}_{2}},\bar{\phi}_{\mathcal{J}_{2}}),
\label{lengths}
\ee
where we defined the $\mathrm{Spin}_0(d,d)$-invariant bilinear form on spinors:
\be
(\phi_1,\phi_2)=(\alpha(\phi_1) \wedge \phi_2)_{\text{top}}.
\ee
Here $\alpha$ reverses the indices of a form:
\be
\alpha(\phi)= \sum_l \phi_{i_l \ldots i_1} dx^{i_1} \wedge \ldots \wedge dx^{i_l}.
\ee
In particular the bilinear form is invariant under the $b$-transform.
To show eq.~\eqref{lengths} we note that for two form $\phi_1$ and $\phi_2$:
\be
\Tr \left( \sla \! \phi_1 \gamma_{1\ldots d} \; \sla \! \phi_2\right) = \frac{1}{\sqrt{g}} \left( \alpha(\phi_1) \wedge \alpha(\phi_2)\right)_{\text{top},\epsilon}.
\label{traceprop}
\ee
To prove this equation we used the fact that all antisymmetrized products of $\gamma$-matrices
are traceless so that the trace selects out the piece proportional to $\mathbf{1}$. Now we plug in $\phi_{\mathcal{J}_1}$ from eq.~\eqref{phiJ}
and use $\left(\eta_R^\dagger \gamma_{i_1\ldots i_l} \eta_L\right)^*=\left(\eta_L^\dagger \gamma_{i_l\ldots i_1} \eta_R\right)$ to calculate $\bar{\phi}_{\mathcal{J}_1}$.
We do the same for $\phi_{\mathcal{J}_2}$. In the end we find simply
\be
c=(-1)^n.
\ee

\subsection{Generalized calibrations in generalized Calabi-Yau manifolds}

With all this machinery we are finally ready to show that the notion of generalized calibrations introduced here is
equivalent to definition 7.10 of \cite{gualtieri}. We start from \eqref{cond2} and introduce a trace over
the spinor representation on the right hand side. This is trivial because the right hand side is a scalar.
Next we use cyclicity:
\be
\sqrt{\det(P(g)+F)} = e^{-i \gamma} \Tr \left(\eta_R^\dagger \rho(F) \eta_L \right)= e^{-i \gamma} \Tr \left(\eta_L \eta_R^\dagger \, \rho(F)\right).
\label{calcalc1}
\ee
Now we can plug in eqs.~\eqref{rhoF3} and \eqref{phiJ} and use eq.~\eqref{traceprop}. We find:
\be
\sqrt{\det(P(g)+F)} = (-1)^t e^{-i \gamma} \sqrt{P(g)} \frac{1}{\sqrt{g}} \left(\phi_{\mathcal{J}_2},\tau_{(N,F)}\right)_\epsilon,
\label{calcalc2}
\ee
where $(-1)^t = (-1)^{(d-p)(d-p-1)/2}(-1)^{p(d-p)}$ is a sign.

The volume form $\mathrm{vol}|_{L(N,F)}$ contains $\sqrt{\det G|_{L(N,F)}}$ where $G|_{L(N,F)}$ is the
pull-back of the metric $G$ to the generalized tangent bundle $L(N,F)$.  We calculate
\be
\sqrt{\det G|_{L(N,F)}} \propto \det(P(g)+F).
\ee
Plugging in eq.~\eqref{calcalc2} we find
\be
\sqrt{\det G|_{L(N,F)}} \propto (\tau_{(N,F)},\phi_{\mathcal{J}_2})(\phi_{\mathcal{J}_2},\tau_{(N,F)}).
\ee
In \cite{gualtieri} the $\mathrm{Cliff}(d,d)$ bispinor $\Omega_2$ is defined as $\Omega_2=(.,\phi_{\mathcal{J}_2})(\phi_{\mathcal{J}_2},.)
\in \wedge^\bullet \left( T_M \oplus T^*_M\right) \otimes \det T^*_M$. Gualtieri argues furthermore that for
a pure spinor $\phi$ the bispinor $(.,\phi)(\phi,.)$ is an element of $\det L \otimes \det T^*_M$ where $L$ is the
corresponding maximal isotropic subbundle. Introducing again a trace, but now over the $(d,d)$-spinors
we use this fact for $\tau_{(N,F)}$ and find
\be
\mathrm{vol}|_{L(N,F)} = e^{i \gamma'} \frac{1}{\sqrt{g}} \Omega_2|_{L(N,F)}.
\label{gencal}
\ee
We introduced the $1/\sqrt{g}$ factor to compensate for the extra $\det T^*_M$ factor in the
transformation law of $\Omega_2$. Eq.~\eqref{gencal} is precisely the definition of the calibration introduced in \cite{gualtieri}.

In the end we find that a D-brane is generalized calibrated if it is a generalized complex submanifold
with respect to $\mathcal{J}_1$ and obeys eq.~\eqref{gencal} for $\mathcal{J}_2$.

\section{Discussion}

In this paper we have introduced generalized calibrations that provide a bound on the Dirac-Born-Infeld
energy, rather than the volume.  We considered a supersymmetric background with non-vanishing dilaton and 3-form
$H$. We showed that
\be
e^{-\Phi} \sqrt{\det(P(g)+F)} \ge \Re \left( e^{-i \gamma} e^{-\Phi} \eta_R^\dagger \rho(F) \eta_L \right),
\ee
with $e^{-i \gamma}$ a constant phase and $\eta_L$ and $\eta_R$ generators of left- and right-moving unbroken supersymmetry. We established
the relation between calibrated submanifolds and supersymmetric cycles. We showed that one obtains the same results
by demanding supersymmetry in the string world-sheet approach. Finally, we made the connection with the calibrations
introduced in \cite{gualtieri} in the context of generalized Calabi-Yau geometry. This latter geometry contains two commuting
generalized complex structures $(\mathcal{J}_1,\mathcal{J}_2)$. We showed that D-branes are calibrated, in the sense of
this paper, if and only if they are complex submanifolds with respect to $\mathcal{J}_1$ and calibrated, in the sense
of \cite{gualtieri}, with respect to $\mathcal{J}_2$, or vice-versa. Furthermore, we note that the mirror map changes
the sign of the U(1) R-symmetry in the right-moving sector and thus sends $J_R \rightarrow -J_R$ and changes $\eta_R$ into
$\eta^c_R$. It also interchanges $\mathcal{J}_1$ and $\mathcal{J}_2$. Thus B- and A-type branes are interchanged.

We would like to emphasize that our analysis only applies to abelian gauge fields. It would be interesting although
presumably extremely hard to generalize to non-abelian gauge fields. In this case the full form of the analogue
of the Dirac-Born-Infeld action is not even known and there is an intricate interplay with derivative corrections.

In this paper we considered $SU(n)_L \times SU(n)_R$ structure. We do not have to restrict ourselves
to this case. In fact, the analysis in section \ref{cal} goes through as soon as we have an $\eta_L$ on the
left and a $\eta_R$ on the right satisfying eqs.~\eqref{susy}. We can then construct a closed calibration, establish
the calibration bound, and show the relation with supersymmetry as before. So we could also consider for example
$Spin(7) \times Spin(7)$ in $d=8$ and $G_2 \times G_2$ in $d=7$. In the latter case we could make the connection
with the generalized $G_2$-structures introduced in \cite{witt,jeschek}. We could define a calibration in the same way
as for a generalized Calabi-Yau geometry. The details remain work in progress.

We saw that for getting supersymmetric D-branes from the world-sheet perspective one also needed to impose
stability as an additional condition in addition to just being topological. However, stability is also important even if one stays
within topological string theory. Indeed, at the quantum level there is an anomaly in the R-charge. For A-branes
without gauge field the anomaly vanishes if the Maslov class vanishes (for a review see section 3.1.1 of \cite{aspinwall}
or sections 38.4 and 39.3 of \cite{mirrorbook}). The definition of the Maslov class is closely related to the notion of ``special''
in special Lagrangian. Specifically, it is easy to show that any special Lagrangian submanifold has vanishing Maslov class.
Although the conditions for anomaly cancellation for the coisotropic A-branes are not known, in \cite{kapustinstable} a proposal
for a generalized Maslov class was made. Even more speculatively, we could also make here
a proposal for a generalized Maslov class for the case $H \neq 0$. Topological branes satisfy eq.~\eqref{cond1} or
rephrased conveniently:
\be
\sqrt{\det(P(g)+F)} = e^{-i \beta(\sigma)} \eta_R^\dagger \rho(F) \eta_L,
\ee
where $e^{-\beta(\sigma)}$ provides a map from the submanifold $N$ on which the D-brane wraps to the circle $S^1$. This in turn
induces a map on the fundamental group $\beta_*:\pi_1(N) \rightarrow \pi_1(S^1) \cong \mathbb{Z}$ which we could take as the generalized
Maslov class. Clearly, if the brane is stable, this Maslov class is trivial.

Another interesting generalization would be to reintroduce the R-R fields in the supersymmetry transformations
\eqref{susy}. The exterior derivative of the calibration will now be related to these R-R fields (see \cite{grana2,grana3}).
It would be nice to introduce a new calibration so that calibrated submanifolds will minimize the sum of the Dirac-Born-Infeld
term and the Wess-Zumino term.

As a final speculation, we note that in \cite{gukov} a relation between calibrations and the effective superpotential was found
in the context of compactification on Calabi-Yau 4-folds. It would be interesting to see if this relation could be
extended to generalized calibrations and generalized Calabi-Yau manifolds and to compare with the results of \cite{grana3}.

\bigskip

\acknowledgments

The author wishes to thank Mark Van Raamsdonk, Alex Sevrin, Greg Van Anders, Moshe Rozali and Gordon Semenoff for
proofreading the manuscript, Jim Bryan for a useful discussion and the theory group at UBC for
positive feedback. This work is supported in part by the Natural Sciences and Engineering Research
Council of Canada.

\appendix

\section{Conventions}
\label{conventions}

We use $\mu,\nu,\ldots$ for space-time indices, $i,j,\ldots$ for indices on the internal manifold $M$ and
$a,b,\ldots$ for indices on the D-brane world-volume. $d$ indicates the dimension of $M$ and $p$ the dimension
of the D-brane. If $d$ even, we define $n=d/2$ and if $p$ even, $k=p/2$.

Given a form $\phi=\frac{1}{l!} \phi_{i_1 \ldots i_l} dx^{i_1} \ldots dx^{i_l}$ we will denote its contraction
with the $epsilon$-tensor as
\be
\phi_{\epsilon} = \frac{1}{l!} \phi_{i_1 \ldots i_l} \epsilon^{i_1 \ldots i_l},
\ee
and the contraction with $\gamma$-matrices as
\be
\sla \! \phi= \frac{1}{l!} \phi_{i_1 \ldots i_l} \gamma^{i_1\ldots i_l}.
\ee
The Hodge dual form is given by
\be
\dual{\phi}^{j_1 \ldots j_{d-l}} = \frac{1}{\sqrt{g}\,l!} \phi_{i_1 \ldots i_l} \epsilon^{i_1 \ldots i_l j_1 \ldots j_{d-l}}.
\ee

$\gamma$-matrices on curved space are as usual defined as $\gamma_i = e^A_i \gamma_A$, where $e^A_i$ is the vielbein and $\gamma_A$
a $\gamma$-matrix on flat space with metric $\delta_{AB}$. $\gamma$-matrices $\gamma_a$ on the D-brane world-volume are defined as the pull-back of
$\gamma$-matrices on the internal manifold: $\gamma_a = \partial_a x^i \gamma_i$. Furthermore we
define
\be
\gamma_{1 \ldots d}=\frac{1}{\sqrt{g}} \gamma_{i_1 \ldots i_d} \epsilon^{i_1 \ldots i_d}.
\ee
Since $\gamma_{1\ldots d}^2=(-1)^{\frac{d(d-1)}{2}}$ we should define the chirality matrix as
\be
\gamma_{d+1}= i^{\frac{d(d-1)}{2}} \gamma_{1 \ldots d}.
\ee

Given a complex structure $J$, the spinor $\eta_0$ --- the {\em empty state} --- is such that it is annihilated by
all $X^i \gamma_i$ where the vector $X$ is a $(+i)$-eigenvalue of $J$. If we choose complex coordinates $\alpha, \beta, \ldots$
such that
\be
J^{\alpha}{}_{\beta} = i \delta^{\alpha}{}_{\beta}, \qquad J^{\bar{\alpha}}{}_{\bar{\beta}} = -i \delta^{\bar{\alpha}}{}_{\bar{\beta}},
\ee
$\eta_0$ is annihilated by all $\gamma_{\alpha}$:
\be
\gamma_{\alpha} \eta_0 =0, \quad \text{for all} \, \alpha
\label{empty}
\ee
The {\em completely filled state} $\eta_0^c$ on the other hand is given by
\be
\eta_0^c = \frac{1}{g^{1/4} 2^{n/2}} \gamma_{\bar{n} \ldots \bar{1}} \eta_0.
\label{filled}
\ee
It satisfies $\gamma_{\bar{\alpha}} \eta^c_0 =0$ for all $\alpha$.

Starting from the complex coordinates we can also define coordinates $x^i,y^i$ in which $J$ takes a block-diagonal form:
\be
z^{\alpha}=\frac{1}{\sqrt{2}} \left( x^{\alpha} + i y^{\alpha}\right), \qquad z^{\bar{\alpha}}=\frac{1}{\sqrt{2}} \left( x^{\alpha} - i y^{\alpha}\right),
\ee
Building the spinor representation by acting with $\gamma_{\bar{\alpha}}$ on $\eta_0$, we find that the $\gamma_{\alpha}$ and $\gamma_{\bar{\alpha}}$
are real in this representation such that $\gamma_{x^{\alpha}}^*=\gamma_{x^{\alpha}}^*$ and $\gamma_{y^{\alpha}}^*=-\gamma_{y^{\alpha}}^*$.
We can then define the charge conjugation matrix $C$ as
\be
\begin{split}
& C=\frac{1}{g^{1/4}} \gamma_{x^n \ldots x^1}, \qquad \text{for} \, n \, {\text{odd}}, \\
& C=\frac{1}{g^{1/4}} (-1)^{n/2} \gamma_{y^n \ldots y^1}, \qquad \text{for} \, n \, {\text{even}}, \\
\end{split}
\label{conj}
\ee
which indeed obeys
\be
\left(\gamma_i\right)^*=C^{-1} \gamma_i C.
\label{conjugation}
\ee
From \eqref{filled} and \eqref{conj} follows
\be
\eta^c=C \eta^*.
\label{upspinor}
\ee
From the defining property \eqref{conjugation} follows that $CC^*=\delta \mathbf{1}$ with
$\delta=\delta^*$.
With our choices $C$ is normalized such that $|\delta|=1$. To be precise
\be
C C^* = (-1)^{\frac{n(n-1)}{2}} \mathbf{1}.
\label{cinvariantsign}
\ee
Note that although we used a specific choice of $\gamma$-matrices and $C$
the remaining sign is independent of that.

\listoftables       
\listoffigures      

\end{document}